\documentclass[10pt, conference]{IEEEtran}
\usepackage{latexsym}
\usepackage{graphicx}
\usepackage{mathptmx}
\usepackage{comment}
\usepackage{amsmath}
\usepackage{amsfonts}
\usepackage{amssymb}
\usepackage{amsbsy}
\usepackage{amsthm}

\usepackage[pdftex,colorlinks=true,urlcolor=blue,citecolor=black,anchorcolor=black,linkcolor=black]{hyperref}

\usepackage{tabularx}
\usepackage{ragged2e}
\newcolumntype{Y}{>{\Centering\arraybackslash}X}
\newcolumntype{Z}[1]{>{\Centering\arraybackslash\hsize=#1\hsize}X}

\usepackage{listings}
\lstdefinestyle{customjava}{
  breaklines=true,
  frame=L,
  xleftmargin=\parindent,
  language=Java,
  showstringspaces=false,
  tabsize=2,
  captionpos=b
}

\begin{document}


\title{Renovation of EdgeCloudSim: An Efficient Discrete-Event Approach}

\author{\IEEEauthorblockN{Raphael Freymann, Junjie Shi, Jian-Jia Chen, and Kuan-Hsun Chen\\}
	\IEEEauthorblockA{Design Automation for Embedded Systems Group\\ Department of Informatics, TU Dortmund University, Germany\\
		\{raphael.freymann, junjie.shi, jian-jia.chen, kuan-hsun.chen\}@tu-dortmund.de\\  
	}
}

\maketitle

\pagestyle{plain}

\begin{abstract}
    Due to the growing popularity of the Internet of Things, edge computing concept has been widely studied to relieve the load on the original cloud and networks while improving the service quality for end-users. To simulate such a complex environment involving edge and cloud computing, EdgeCloudSim has been widely adopted. However, it suffers from certain efficiency and scalability issues due to the ignorance of the deficiency in the originally adopted data structures and maintenance strategies. Specifically, it generates all events at beginning of the simulation and stores unnecessary historical information, both result in unnecessarily high complexity for search operations. In this work, by fixing the mismatches on the concept of discrete-event simulation, we propose enhancement of EdgeCloudSim which improves not only the runtime efficiency of simulation, but also the flexibility and scalability. Through extensive experiments with statistical methods, we show that the enhancement does not affect the expressiveness of simulations while obtaining 2 orders of magnitude speedup, especially when the device count is large. 
\end{abstract}

\begin{IEEEkeywords}
Edge Computing, Cloud Computing, Discrete-Event Simulation
\end{IEEEkeywords}

\section{Introduction}
\thispagestyle{plain}
\label{sec:intro}
Nowadays edge computing as a new computing paradigm has attracted more and more attention. With the proliferation of the Internet of Things (IoT) and the stagnation of network bandwidth development, the original design of cloud computing is no longer sufficient \cite{EdgeComp}. The number of devices in the network has increased while the devices are acting as not only data consumers but also producers, e.g., computer vision tasks are deployed in resource-constrained edges~\cite{DBLP:conf/ccgrid/TomaWLC19}. 


Specifically, such devices may offer computing and/or network resources as well~\cite{ECS1}. It therefore makes sense to process this data in the geographic vicinity of the devices, rather than uploading it to the cloud and waiting for the results to come back. However, in such scenarios the network is often considerably crowded and the edge devices are also mobile, so that response time of tasks might not be stable, which might greatly affect the end-user experience. In addition, limited resources on edge devices, e.g., energy and computational power are also challenges~\cite{8449105}. In order to study various upcoming challenges, simulation environments are often preferred, since deploying a case study in practice is too disruptive~\cite{Law2000}.
\begin{figure}[ht]
    \begin{center}
        \includegraphics[width = 1\linewidth]{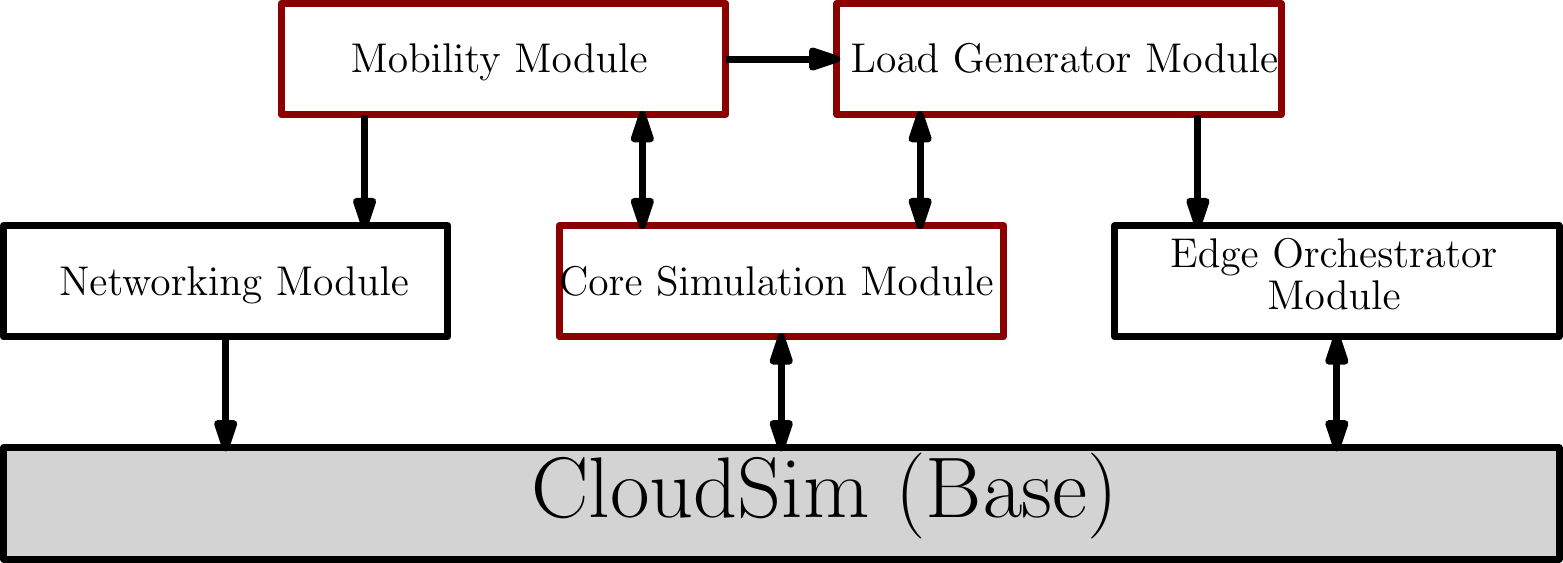}
        \caption{Overview of EdgeCloudSim: Blocks with red boarders are enhanced modules.}
        \label{ECSCD}
    \end{center}
\end{figure}

EdgeCloudSim~\cite{ECS2} is a widely-adopted simulation environment, that is developed to model such edge computing scenarios including network and computational models. Many prominent techniques, e.g.,~\cite{8449105},~\cite{CASADEI2019154}, and~\cite{app8071160}, evaluate the performance of their approaches based on this simulation environment. Similar to iFogSim~\cite{iFogSim}, EdgeCloudSim also relies on CloudSim~\cite{CS:2010}, which is a discrete-event simulator that enables modeling and simulation of cloud systems and application provisioning environments. However, the additional modules introduced by EdgeCloudSim in fact confront certain efficiency and scalability issues. For instance, as proposed by Law and Kelton~\cite{Law2000}, the simulated system should only change its state when an event occurs. However, this is exactly not a case in EdgeCloudSim. As the locations of simulated devices at each time point is determined before the start of simulation, they in fact changes their locations between events in the simulation. Hence, the searching process over a data structure has to be triggered at each time the device's location is needed,
even if the location has not been changed since the last time. Besides, there are many unnecessary operations required before and during the simulation that can greatly degrade the scalability of EdgeCloudSim.

\vspace{0.1in}
\noindent\textbf{Our Contributions:} In this work, we focus on explaining the potential issues that arise from the current design of EdgeCloudSim  and provide a comprehensive extension to overcome these issues.
\begin{itemize}
    \item We discuss two major issues in the original design of EdgeCloudSim~\cite{ECS2}, namely the state of the system with respect to the location of devices changes between the occurrence of two events, and that all simulated events are generated beforehand and added into the event queue offline (see Section~\ref{sec:design-of-edgecloudsim}).
    \item The corresponding enhancement is presented for the aforementioned two major issues without scarifying the accuracy of the simulation models, while the other relevant components in EdgeCloudSim are also refined accordingly (see Section~\ref{sec:enhan}).
    \item We conduct extensive experiments over different configurations to show that the enhanced design can obtain in general two orders-of-magnitude speedup. With two statistical methods, we argue that the enhancement does not affect the expressiveness of simulations (see Section~\ref{sec:eval}). Detailed results and the corresponding scripts are publicly available on \cite{EdgeCloudSim-imp}.

\end{itemize}

\section{Design in EdgeCloudSim}
\label{sec:design-of-edgecloudsim}

In this section, we first overview the original design in EdgeCloudsim, where CloudSim is used to bring the basic functionalities for the modeling of the cloud systems, and is responsible for the general execution of the simulation. In this work, we only enhance some modules introduced by EdgeCloudSim without changing any design of CloudSim. For further information of CloudSim, please refer to \cite{CS:2010}.
Figure~\ref{ECSCD} illustrates an overview of the modules in EdgeCloudSim, details are as follows:

\begin{itemize}
    
    \item \textbf{Core Simulation Module} is responsible for reading the configuration files and setting up the simulation accordingly. EdgeCloudSim allows to set the configuration of the data centers, the properties of the applications, 
    and other basic settings. In addition, it records the results of the simulation by using log file.
    
    \item \textbf{Edge Orchestrator Module} is applied for managing the resources of the system in order to improve the performance of the system. 
    The \textit{Edge Orchestraotr} can start or stop virtual machines and manage the computational resources of hosts. It also decides the location where to assign tasks, i.e., on edge servers or cloud servers, by coordinating these two kinds of servers.
    
    \item \textbf{Networking Module} is responsible for determining the transmission delay of download and upload in wide area network and wireless local area network. The network connection quality changes according to the number of devices in the corresponding sphere, the location of devices, and the workload. Therefore the transmission delay between two entities is dynamic.
    
    \item \textbf{Mobility Module} is responsible for modeling the mobility of devices in systems. 
    In a real system, the movement of a device can cause disconnection, when it is out of the maximum transmission distance, so influences the overall performance of the system. 
    In addition, mobility can cause congestion at the access points. 
    
    \item \textbf{Load Generator Module} is responsible for generating the tasks. Each device 
    is in one of the exclusive period, i.e., the active or inactive period. In the active period devices generate tasks by following a given distribution. In the inactive period, devices do not generate tasks.
    
\end{itemize}
In the following, we detail the design of two modules specifically, i.e., Mobility Module and Load Generator Module, which are the main bottleneck of the runtime efficiency. Afterwards, we clarify the tackled problems in this work. 

\subsection{Mobility Module}
Mobility in EdgeCloudSim is modeled by the \emph{nomadic mobility model}.
That is, each access point has a certain attractiveness. The attractiveness of the access point determines the duration of a device stays in its sphere of influence. 
The value of attractiveness equals to the average duration of a device stays at this access point. Each device randomly chooses an access point (location) to stay at, where all locations have the same probability to be chosen. 
Then the waiting time for a device at its location is drawn from the exponential distribution with the corresponding attractiveness of the location as the expected value. 
When the duration runs out, the device randomly chooses a new location. The above process is used to determine the location for each mobile device in the system.

In the implementation of EdgeCloudSim, the information for locations of each device is stored in a list of trees, that the timestamps of the movement is the index and the location is the value. Each location gets its own random exponential distribution generator, that the attractiveness of that location is the expected value. Then the movements are calculated for each device. 
In each movement, as long as the timestamp of the last movement is smaller than the simulation duration, a random new location is chosen. 
Depending on the attractiveness, the waiting time at this location is generated by using the corresponding random generator. The tuple of time and destination of the movement is put into a tree, where the time is the index and the location is the value. 

Aforementioned processes are repeated until the timestamp of the last movement is greater than the simulation duration. In the end, each device has a tree that contains the destinations of the movements and the time when the movement happens. For each device such a tree is generated and stored in a list before the start of the simulation. 
If the location of a device at a certain time needs to be determined during the simulation, the corresponding tree has to be searched. The corresponding value with the largest index that is smaller than the specified time is the current location of the device.

\subsection{Load Generator Module}
The activity of devices in EdgeCloudSim is modeled by an \emph{idle active load} model, where each device has a task type that it can generate, such as \textit{health app}. 
When a device is in an active period, it can generate tasks according to the given type. The times at which tasks are generated are defined by a Poisson arrival process. 
Each task type has a certain expected value. The time interval in which two tasks of a device are generated follows a random process by the exponential distribution with this expected value. The lengths of the active and inactive periods are fixed and determined by the task type.
Once a task is generated, all the properties of this task are formed in an event, which is put into the future event queue of the simulation. 
The entry time of this event is the time point when the task is generated.
Such generation of tasks is repeated in each period until the end of the simulation or the simulation reaches the maximum defined threshold.
Once the scheduled event occurs, the task is created according to the generated properties. The generated task is further processed by the Mobile Device Manager. 
In this implementation, the data for all tasks is generated before the start of simulation. However, the task is only created at the time of its occurrence.

\subsection{Problem Definitions}
We discover two major issues in the original design affecting the required execution time:
\begin{itemize}
    \item \textbf{Mobility Module} does not take advantage of the nature of a discrete-event simulation. In discrete-event simulation, the state of the system should not change between the occurrence of two events~\cite{Law2000}. However, in the original EdgeCloudSim, devices may change their location between events, since their locations at each time point are already determined before the simulation. Hence, it is necessary to search the location tree of a device, although it is possible that the location has not changed since the last check, i.e., the system does not know when the location of a device changes. 
    \item \textbf{Load Generator Module} generates all properties of tasks before the start of simulation and schedules their generations by adding them into the event queue. When a large system needs to be simulated, it can be several thousands or even millions of events that are created and added to the event queue in the beginning, which is actually not necessary. The sorting of the huge event queue can lead to several problems, e.g., take a large amount of time or out of memory.
\end{itemize}
In addition, \emph{MobileDeviceManager} in Core Simulation Module poses a potential issue when the number of tasks is large.
As long as a task is created, it is bound to its corresponding device. The binding process, i.e., \emph{getById} method, searches over a list of tasks iteratively. However this trivial searching process may become inefficient when the simulation runs longer, since most information in the list is redundant. 



\section{Enhancement}
\label{sec:enhan}

This section explains the proposed enhancements in detail. 
The objective of this work is to improve the existing implementation of EdgeCloudSim with respect to its computation efficiency without affecting the accuracy of the results. To achieve this, the implementation of 
computing Mobility Module and Load Generator Module are enhanced, without modifying the underlying simulation models. Furthermore, one class in Core Simulation Module and the other relevant components are modified in order to meet the functionality of the enhanced two modules.

\subsection{Mobility Module}
\label{MobilityModel}
The objective to improve the implementation of the Mobility Module is to realize event-based location changes,
and hence reduce the overhead to check the trees for the information of movement and locations during the simulation.

In the original design, each device generates a tree to store the information of movement and locations. 
The network module using the information of devices located at the respective access point each time to determine the upload and download delays. Therefore, the locations for all the devices are searched according to their trees for the current entry. 

One possible method to reduce the overhead is to introduce movement events to the simulation. Instead of constructing a tree for each device to store all the information of movement and locations, a device can move dynamically during the simulation when a movement event occurs.
This allows the simulator to store the number of devices at each location, and the number keeps the same between two events. 
And the number only needs to be modified when a movement event occurs for a device. 
The advantage of this method is that these values can be stored in a single array, whose values can be directly retrieved without searching. 
In addition, in the original design of EdgeCloudSim, the settings document is parsed at each computed movement to obtain the location data of the next access point. It is more efficient to store the data for future movements.



{\footnotesize
\begin{lstlisting}[language=Java, firstline=73, lastline=89, numbers = left, caption = {Movement of a device}, label = {IMM}, style = customjava, basicstyle=\scriptsize]
public void move(int deviceId){
    boolean placeFound = false;
    int currentLocationId = deviceLocations[deviceId].getServingWlanId();

    while(placeFound == false){
        int newDatacenterId = SimUtils.getRandomNumber(0,SimSettings.getInstance().getNumOfEdgeDatacenters()-1);
        if(newDatacenterId != currentLocationId){
            placeFound = true;
            --datacenterDeviceCount[currentLocationId];
            ++datacenterDeviceCount[newDatacenterId];
            deviceLocations[deviceId] = datacenters[newDatacenterId];
            double waitingTime = expRngList[newDatacenterId].sample();
            SimManager x = SimManager.getInstance();
            x.schedule(x.getId(),waitingTime,SimManager.getMoveDevice(), deviceId);
        }
    }
}
\end{lstlisting}

}
The detailed implementation is shown in Listing~\ref{IMM}.
First, a new access point is selected towards which the device will randomly move. When this location is found, in the array that stores the number of devices at an access point, the value for the old location is decreased by one. The value for the new access point is incremented by one, and in the array that stores the locations of the devices, the location for the moved device is updated. Then the wait time for the device at the new location is generated using the same model as in the original design. Finally, the next movement of the device is scheduled as an event that occurs after the waiting time expires. The initialization method was also modified according to this principle. 
To achieve this, two new functions were added, namely \textit{readDatacenters} scans and stores the location data of each data center to speed up the accesses, and \textit{getDeviceCount} is used to directly retrieve the number of devices at an access point without collecting the location of all devices.

\subsection{Load Generator Module}
\label{LoadGenerator}

The objective to improve the implementation of the Load Generator Module is to reduce the number of events in the event queue, in order to reduce the overhead of sorting the queue and inserting new element(s) into the queue for large simulation scenarios. 

In the original design, 
the generations of 
all task properties are at the beginning of the simulation.
However, the creation of a task can only take place when it is to be sent from a device. 
All the information for future tasks is stay unused during the run time of the simulator.
Therefore, the load generator is modified so that only the task properties for the next active period of a device are generated and inserted to the event queue. 
Towards this, another type of event is introduced to control the generation of task properties, i.e., to schedule the generation of the task properties for the next active period.

{\footnotesize
\begin{lstlisting}[language=Java, firstline=101, lastline=112, numbers = left, caption = {Generation of TaskProperties}, label = {GTP}, basicstyle=\scriptsize, style = customjava]
public void createTask(int deviceId){
    SimManager sm = SimManager.getInstance();
    double virtualTime = taskRng[deviceId].sample();
    double clock = CloudSim.clock();

    while(virtualTime < activePeriods[deviceId]) {
        sm.schedule(sm.getId(), virtualTime, sm.getCreateTask(), new TaskProperty(deviceId,taskTypeOfDevices[deviceId], clock + virtualTime, expRngList));
        double interval = taskRng[deviceId].sample();
        virtualTime += interval;
    }
    sm.schedule(sm.getId(), activePeriods[deviceId] + idlePeriods[deviceId], sm.getGenTasks(), deviceId);
}
\end{lstlisting}
}

The detailed implementation is shown in Listing~\ref{GTP}. As long as the virtual time is smaller than the duration of the active Period, a new task is generated and scheduled. Then, the virtual time is increased by a random time interval generated using the same model as in the original design. This happens until the virtual time exceeds the active period. At the end, a new event is scheduled, and aforementioned behavior is repeated for the next active period.

\subsection{Mobile Device Manager}
\label{MobDevMan}
\textit{MobileDeviceManager} in Cloud Simulation Module poses a potential efficiency issue because of the iterative search process over a list of tasks \textit{CloudletList} for binding generated tasks on simulated devices. The task in EdgeCloudSim is an extension of the \textit{Cloudlet} in CloudSim, and the \textit{MobileDeviceManager} in EdgeCloudSim extends the \textit{DatacenterBroker} class of Cloudsim. Every time when a task is generated, it is appended at the end of the list and the binding process \emph{bindCloudletToVm} is triggered and call \emph{getById} function.

\begin{lstlisting}[language=Java, numbers = left, caption = {Retrieving a Cloudlet from \textit{CloudletList}}, label = {getid}, style = customjava, basicstyle=\scriptsize]
public static <T extends Cloudlet> T getById(List<T> cloudletList, int id) {
		for (T cloudlet : cloudletList) {
			if (cloudlet.getCloudletId() == id) {
				return cloudlet;
			}
		}
		return null;
}

public void bindCloudletToVm(int cloudletId, int vmId) {
	CloudletList.getById(getCloudletList(), cloudletId).setVmId(vmId);
}
\end{lstlisting}




The underlying implementation of \textit{getById} function (see Listing~\ref{getid}) in fact searches over the list of CloudLet (tasks) to find the matching id. This function is activated every time after the generation of a task. However tasks are never removed from the list even after they are completed. As a result, the size of the list keeps increasing, with more and more redundant historical information that decreases the efficiency of search operations applied to the list. 




Since \textit{SimLogger} is used to output the results of the simulation, which has already stored all the relevant information of tasks, there is no need to keep all the tasks in the list any more.
\textit{MobileDeviceManager} only needs to store the information of tasks that are currently executed in the simulated system. 
All tasks that have finished their execution can be safely removed. As a result, the size of \textit{CloudletList} is significantly reduced and the time for searching a task in the list stays steadily short.

\subsection{Further Changes}
\label{FurtherChanges}

To integrate the enhanced modules, the central class that controls the simulation, i.e., \textit{SimManager} has to be refined. Three new event types are included: 1) the event that is used to trigger the movement of a device; 2) the event that is used to trigger the generation of tasks for an active period; and 3) the event that is used to trigger \textit{SimLogger} to log the current location data. 
Along with the introduced events, the corresponding handlers should be adjusted as well. For example, as only the current location of a device is stored now, \textit{SimLogger} should log the information of location on the fly instead of at the end of simulation.
In the end, Network Module is modified, since it no longer iterates over all devices to determine the number of devices in a location. It now utilizes the new functionality of the Mobility Module to get the number of devices at an access point directly.

\section{Validation and Evaluation}
\label{sec:eval}
In this section, we extensively validate and evaluate the enhanced design. To this ends, two statistical methods are adopted, i.e., the Kolmogorov-Smirnov test (KST) \cite{encstat} and the Q-Q plot \cite{QQ}, to support the validation process with statistical arguments. 
Afterwards, we compare the performance between the original design and the enhanced design, simply based on the spent elapsed time of the simulation runs. If the performance of EdgeCloudSim has been improved, the average execution time of enhanced design should be significantly lower than that of the original design for the same scenarios.
\subsection{Kolmogorov-Smirnov Test and Q-Q Plot}
The KST reports two values, the statistic D and the p-value, where D is the maximum vertical distance between the empirical cumulative distribution functions of the two samples over the original and enhanced designs. This statistic is compared to critical values of the Kolmogorov distribution, and if it is greater than the critical value, the \emph{null hypothesis} that both samples come from the same distribution is rejected. The p-value is the probability of obtaining results that are at least as extreme as the observed one under the assumption that the null hypothesis is true. 
Since the null hypothesis is a sufficient test to support our arguments, we also provide Q-Q plots to gain more insights.


\begin{table*}[h]
  \centering
    
    \footnotesize
    \begin{tabularx}{\textwidth}{|c|Y|Y|Y|Y|Y|Y|}
      \hline
      &\multicolumn{2}{Z{2}|}{Single Tier}& \multicolumn{2}{Z{2}|}{Two-Tier} & \multicolumn{2}{Z{2}|}{Two-Tier with Orchestrator}\\
      \hline
      Metric & statistic D & p-value & statistic D & p-value & statistic D & p-value\\
      \hline
      \# of Tasks Generated&0.036&0.9022&0.060&0.3291&0.040&0.8186 \\
      \hline
      Failed (Rel)&0.044&0.7184&0.034&0.9347&0.116&0.0024\\
      \hline
      Avg Serv. Time&0.028&0.9895&0.048&0.6121&0.050&0.5596 \\
      \hline
      Failed (Mob)&0.068&0.1979&0.084&0.0587&0.078&0.0955\\
      \hline
      Failed (VM)&0.038&0.8632&0.034&0.9347&0.006&1.0000\\
      \hline
    \end{tabularx}
    \vspace{0.1in}
    \caption{KST results of sample application 1 over three architectures with 500 devices for 30 minutes. \label{KST} }
\end{table*}

\subsection{Evaluation Setup}
\label{Data}
In order to collect the relevant data for the comparison, the sample applications given by EdgeCloudSim are adopted. 
Since several dynamic influences are simulated by probability distributions in EdgeCloudSim, we conducted 500 runs for each comparison in order to obtain sufficiently large samples against randomness. To investigate if and how device counts and simulation duration have an impact on the comparability of the results, the scenarios are executed with varying device counts and simulation duration parameters.
Other parameters of the sample applications are not changed at all. All required data is provided by EdgeCloudSim originally, so no further measures need to be taken in this regard.

For time measurement, the same applications are also adopted and the file logging was deactivated for both versions for the time measurement.. They executed on an Ubuntu 20.04.2 PC with an i5-8300H CPU with Turbo-Boost disabled and with no Hyper-Threading at a base clock of 2.30 GHz and with 16 GB memory. To analyze the impact of different numbers of devices, 200, 400, 600, 800, and 1000 devices are simulated for a simulation duration of 30 minutes. To study the performance for different simulation duration, 200 devices are simulated for simulation duration of 30, 60, 90, 120, and 150 minutes. The average execution times of 30 iterations for each scenario are compared. 
This sample size should be sufficient to detect a significant difference between the original simulator and the enhanced one. 
The required execution time is measured in seconds. Since individual runs have duration of a few seconds to a few minutes, time differences of milliseconds or smaller are insignificant.

\begin{figure}[t!]
    \begin{center}
            \includegraphics[width = \linewidth]{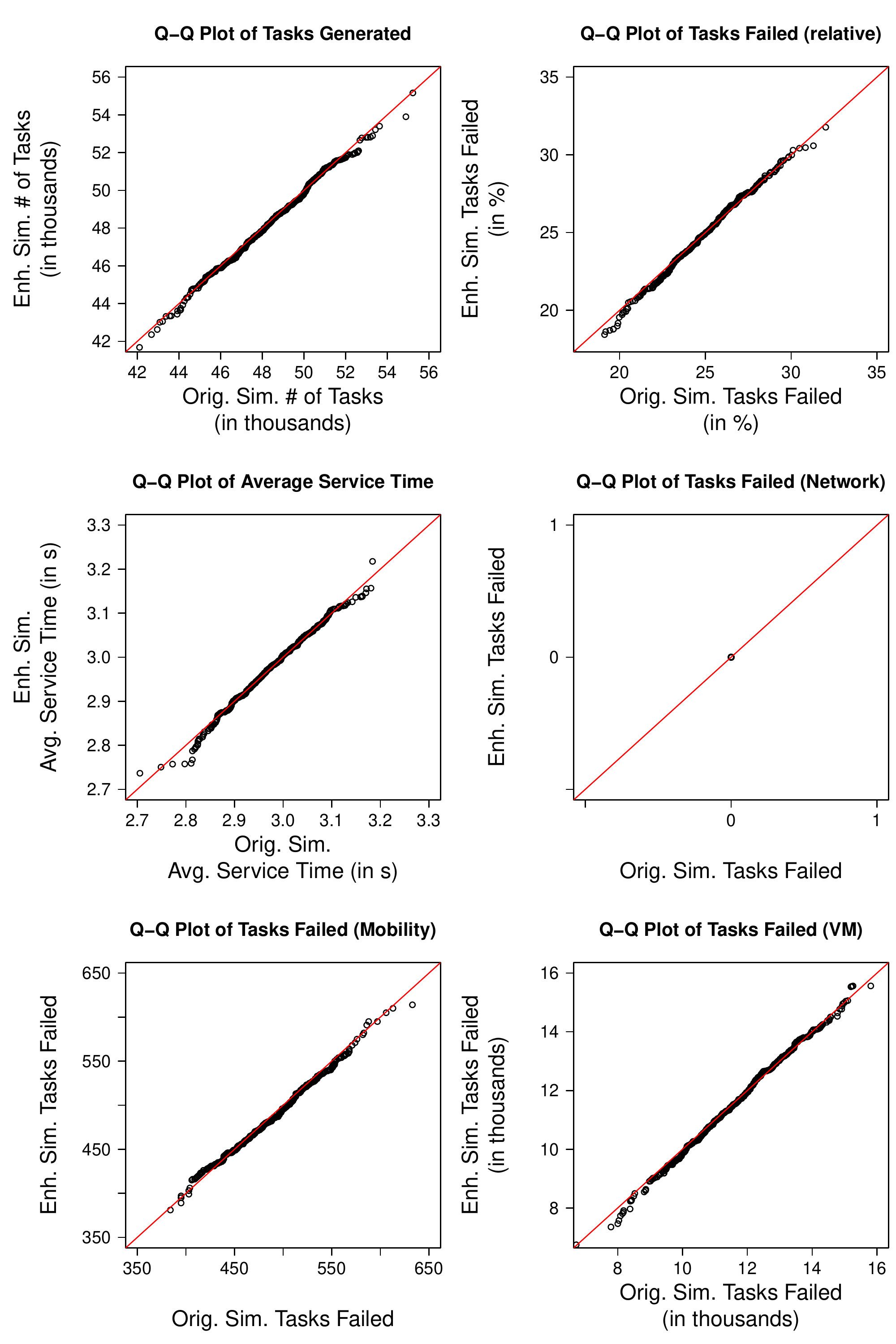}
            \caption{Single tier architecture: 500 devices, 30 minutes and sample app 1}
            \label{QQS1ST}
    \end{center}
\end{figure}

\subsection{Validation of Compatibility}
\label{CompRes}
To valid the enhancement, we examine three built-in applications on three different architectures: 1) single tier, 2) two-tier, and 3) two-tier architectures with edge orchestrator.
We mainly focus on the first sample application provided by the original simulator, as the second and third sample applications result in similar trends, which do not provide additional insights. For further details, the extensive results can be found in the repository~\cite{EdgeCloudSim-imp}. 
For the load generator module, we compare the total number of generated tasks, the failure rate of the simulated architecture, i.e. the percentage of failed tasks, and the average service time, i.e. the average time that elapsed between sending a task from a device and the arrival of the result.
For the mobility module, we take a closer look at the types of failures. That is, the number of failures caused by the network, the movement and the load of the virtual machines. The other results of the simulation are not examined specifically, since they have a strong correlation with the values examined. 

For all architectures, we compare the original and enhanced designs in terms of the number of tasks created, the
ratio of failed tasks, the average service time as the most important simulation results, and the individual failure types. We compare the individual failure types
in more detail, in order to explore whether the enhancement of mobility module 
leads to the shifts among individual failure types. In the following evaluation, 500 devices are simulated. The simulation duration is 30 minutes. 500 iterations are executed. As significance level $\alpha$ for the KST 0.05 is chosen. This also applies to all future tests. In each Q-Q plot, the results of original design is plotted as a red line with slope 1 that goes through the origin.
\begin{figure}[t]
    \begin{center}
        \includegraphics[width = \linewidth]{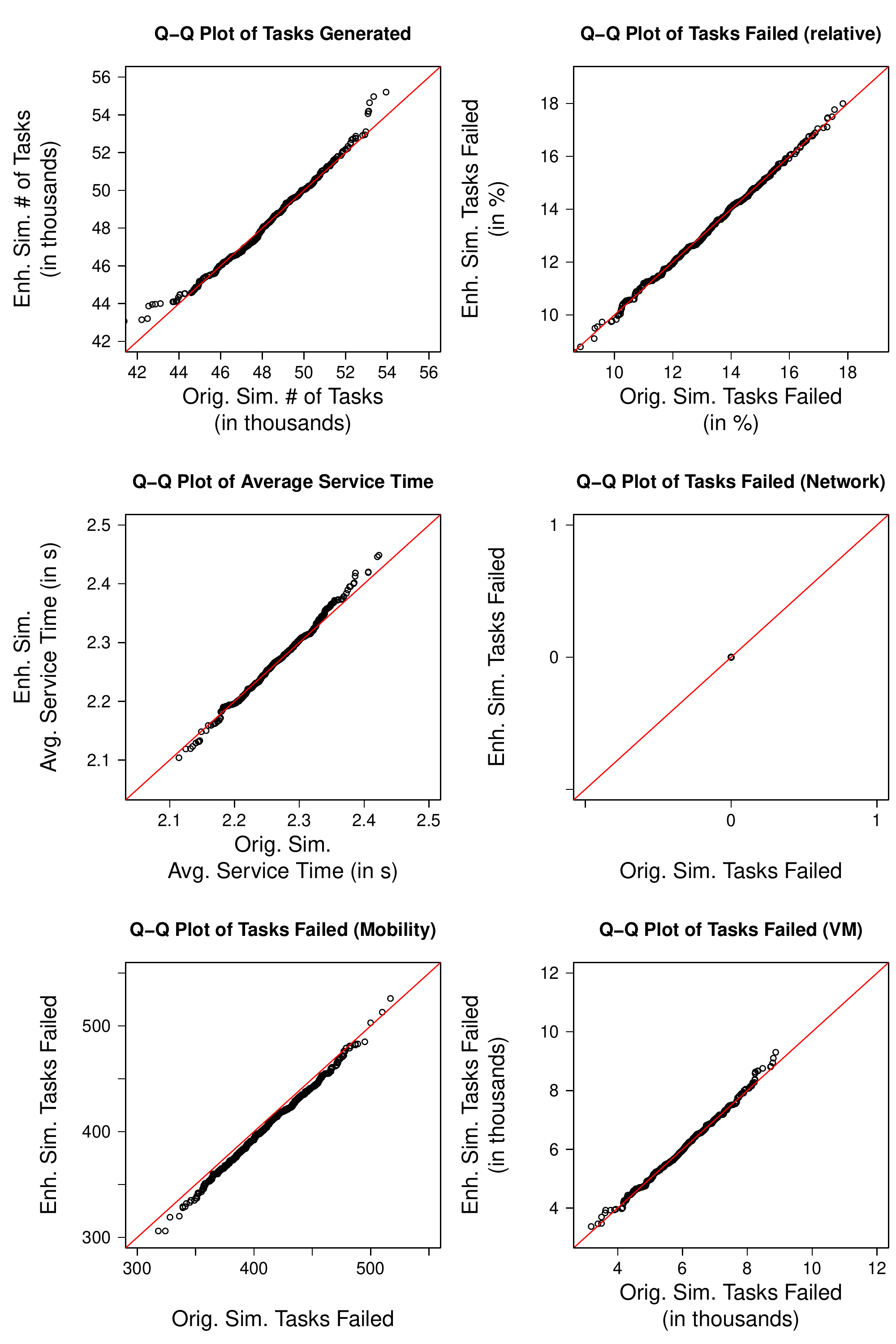}
            \caption{Two tier architecture: 500 devices, 30 minutes and sample app 1}
            \label{QQS1TT}
    \end{center}
\end{figure}

\begin{figure}[h!]
        \begin{center}
                \includegraphics[width = \linewidth]{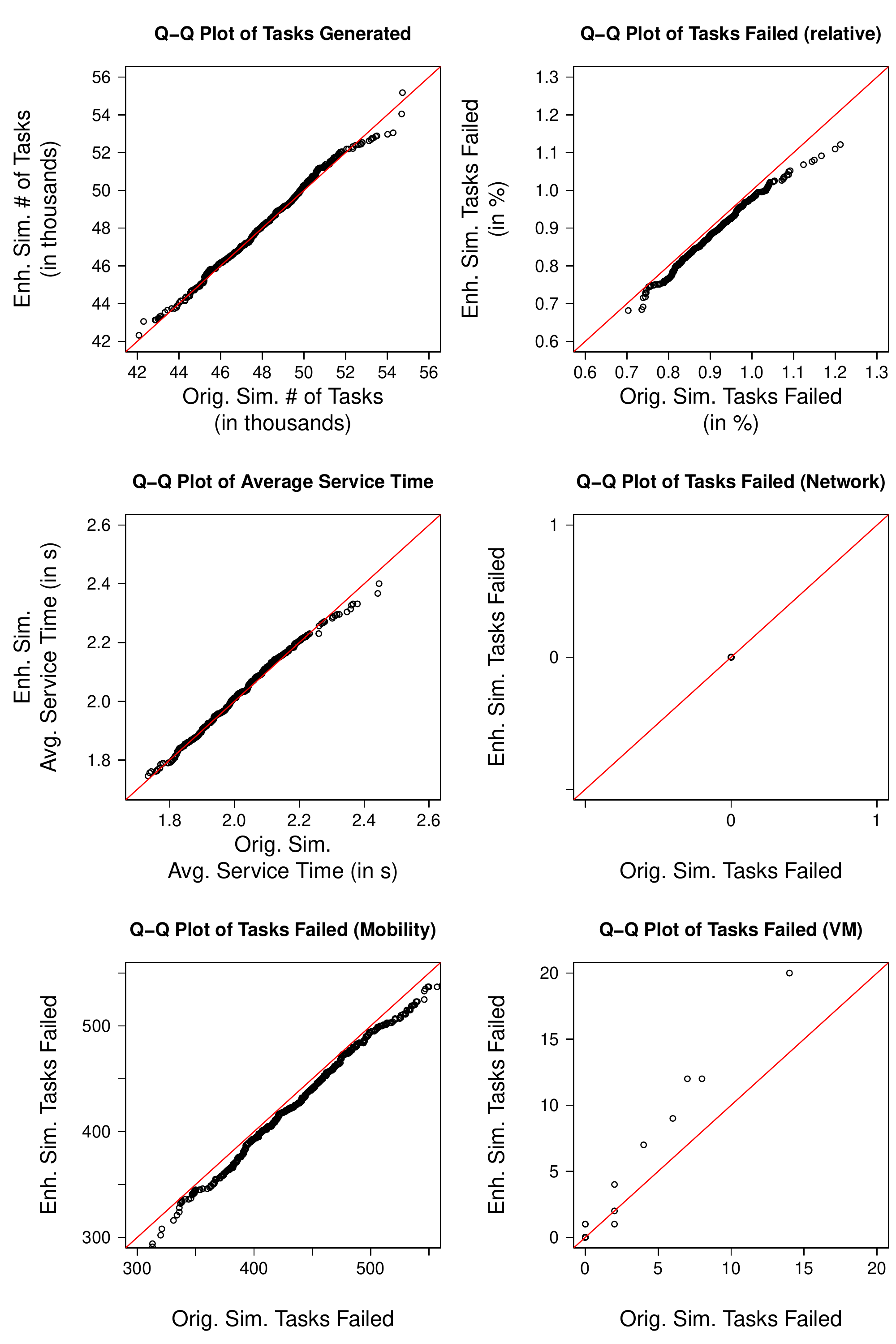}
                \caption{Two tier architecture with an edge orchestrator: 500 devices, 30 minutes and sample app 1}
                \label{QQS1TTEO}
        \end{center}
\end{figure}

Figure \ref{QQS1ST} shows the Q-Q plots for the single tier architecture. For all metrics the plots follow a line that goes through the origin. From the associated results in Table~\ref{KST}, the null hypothesis of identical distribution cannot be rejected, i.e., none of p-value is less than 0.05. No task failed due to network problems in both cases.
Figure~\ref{QQS1TT} shows the corresponding Q-Q plots for the two-tier architecture. We can see from the Q-Q plot of tasks failed due to mobility failures that there is a slight shift. Slightly fewer tasks fail in the enhanced design, but this difference is still not statistically significant, as shown by the Table~\ref{KST}.
Figure \ref{QQS1TTEO} shows the corresponding Q-Q plots for the two-tier architecture with edge orchastrator. The enhanced design leads to a slightly reduced number of mobility failures. However, it can be seen from the Q-Q plots that this effect is negligible.
The only factor that has an impact on the mobility failures is the average service time. The longer a task takes to be processed and to be returned, the higher the chance that the device has moved in between. However, this factor is also one of the results of the simulation and there is no difference between two designs in this respect. 
In Table~\ref{KST}, the KST does not reject the null hypothesis for the number of mobility failures, but rejects the hypothesis for the rate of total failed tasks. However, almost no other failure types occurred in this examination, so the mobility failures dominate the rate of overall failed tasks leading to the rejection. From the above observation in the Q-Q plots, we can only observe a small difference between the designs regarding the number of mobility failures. 




 

Overall, no significant difference between the results of the original simulator and the enhanced one is noticeable. 
Due to the original design of EdgeCloudSim, the simulation environment was not deterministic already. Hence, it is impossible to derive the exactly same results, even without applying the introduced enhancement. Along with the above results, we can identify that the differences between the counts of mobility failures are negligible.  Hence, we conclude that the expressiveness of EdgeCloudSim is not affected.

\subsection{Required Execution Time}
\label{PerfEval}

\begin{figure}[h!]
        \begin{center}
                \includegraphics[width = \linewidth]{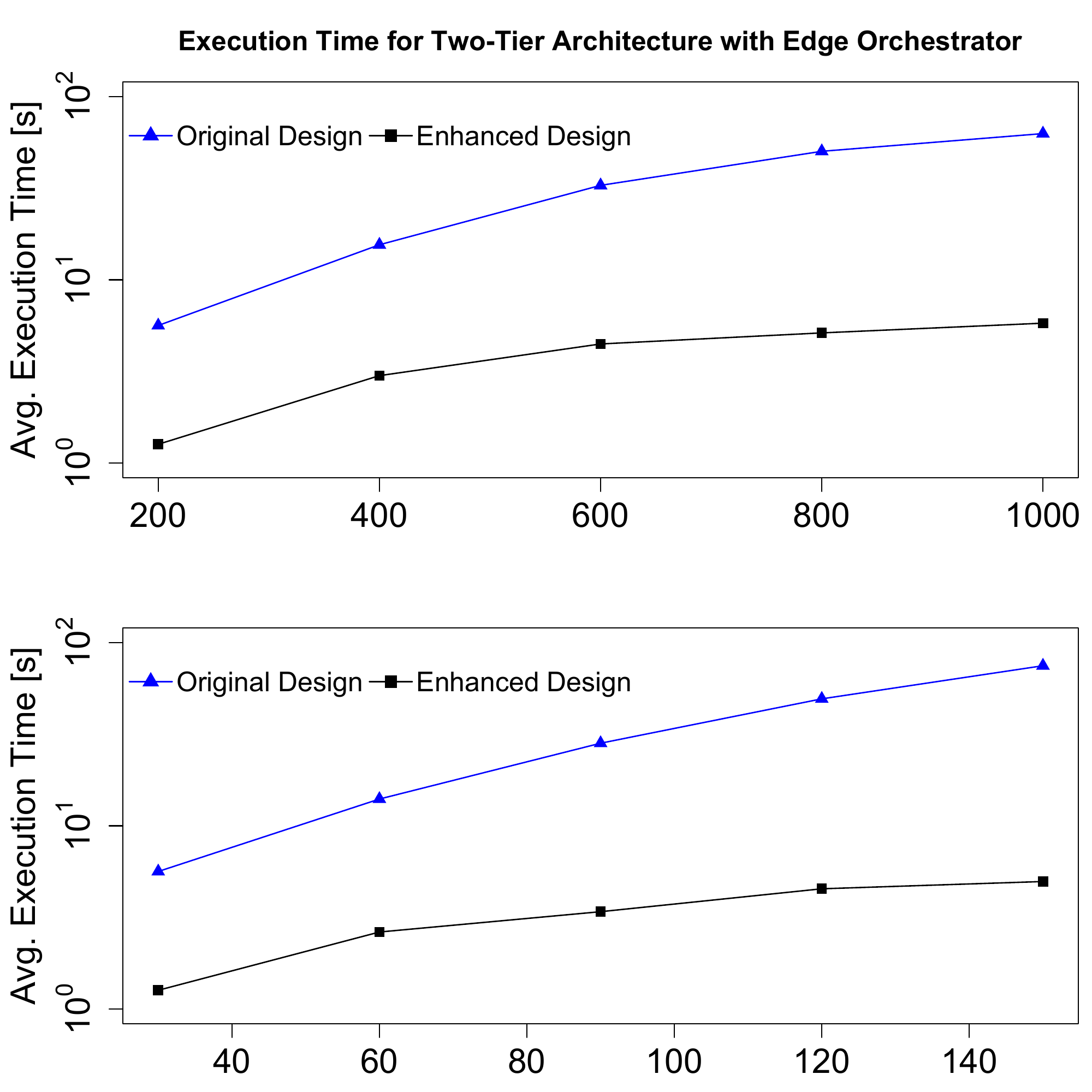}
                \caption{Execution time for two-tier with edge orchestrator scenario of sample app 1 with varying device count and duration (Y-axis is in log-scale).}
                \label{TGS1TTEO}
        \end{center}
\end{figure}
\begin{figure}[h!]
        \begin{center}
                \includegraphics[width = \linewidth]{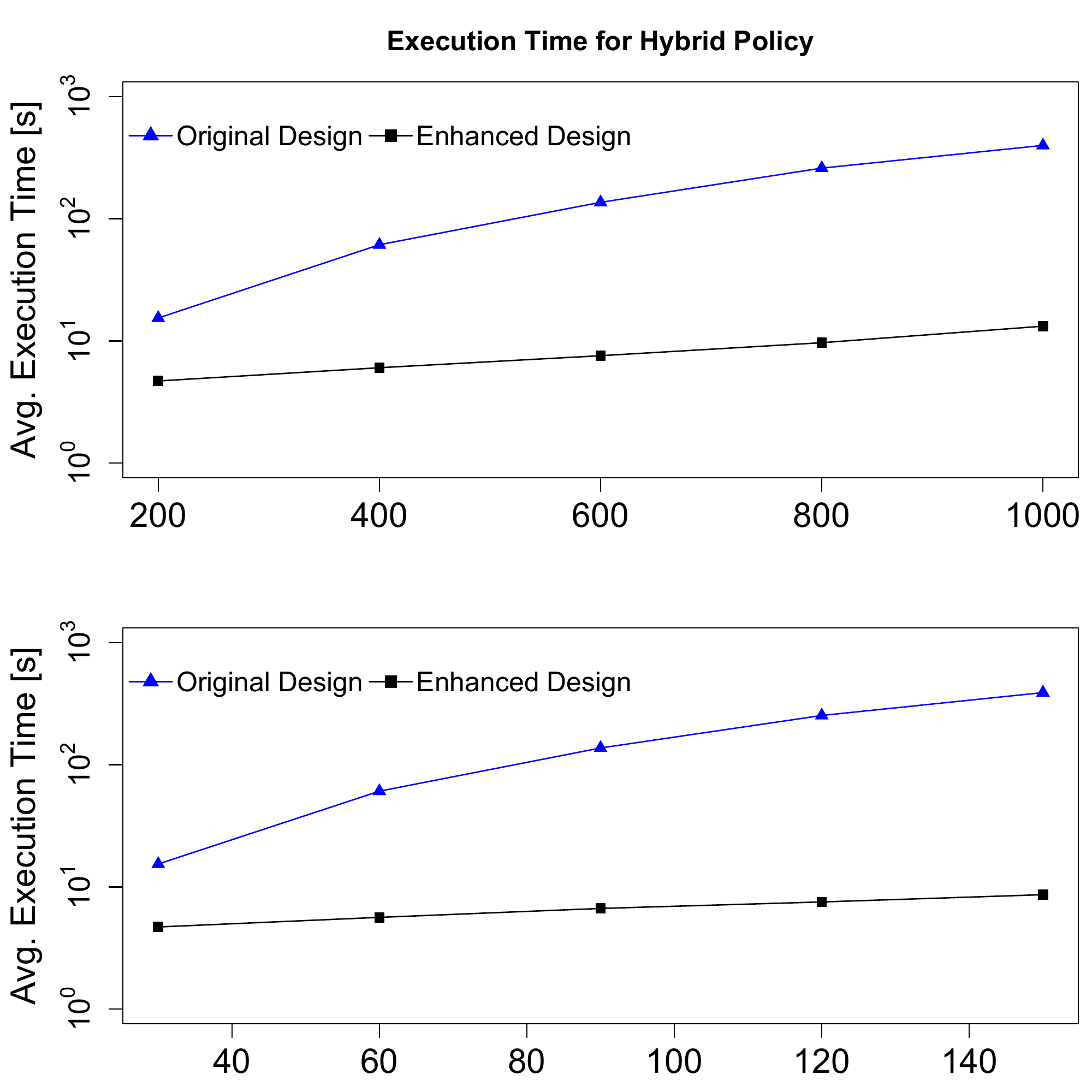}
                \caption{Execution time for hybrid policy of sample app 2 (Y-axis is in log-scale).}
                \label{TGS2HY}
        \end{center}
\end{figure}
\begin{figure}[t]
        \begin{center}
                \includegraphics[width = \linewidth]{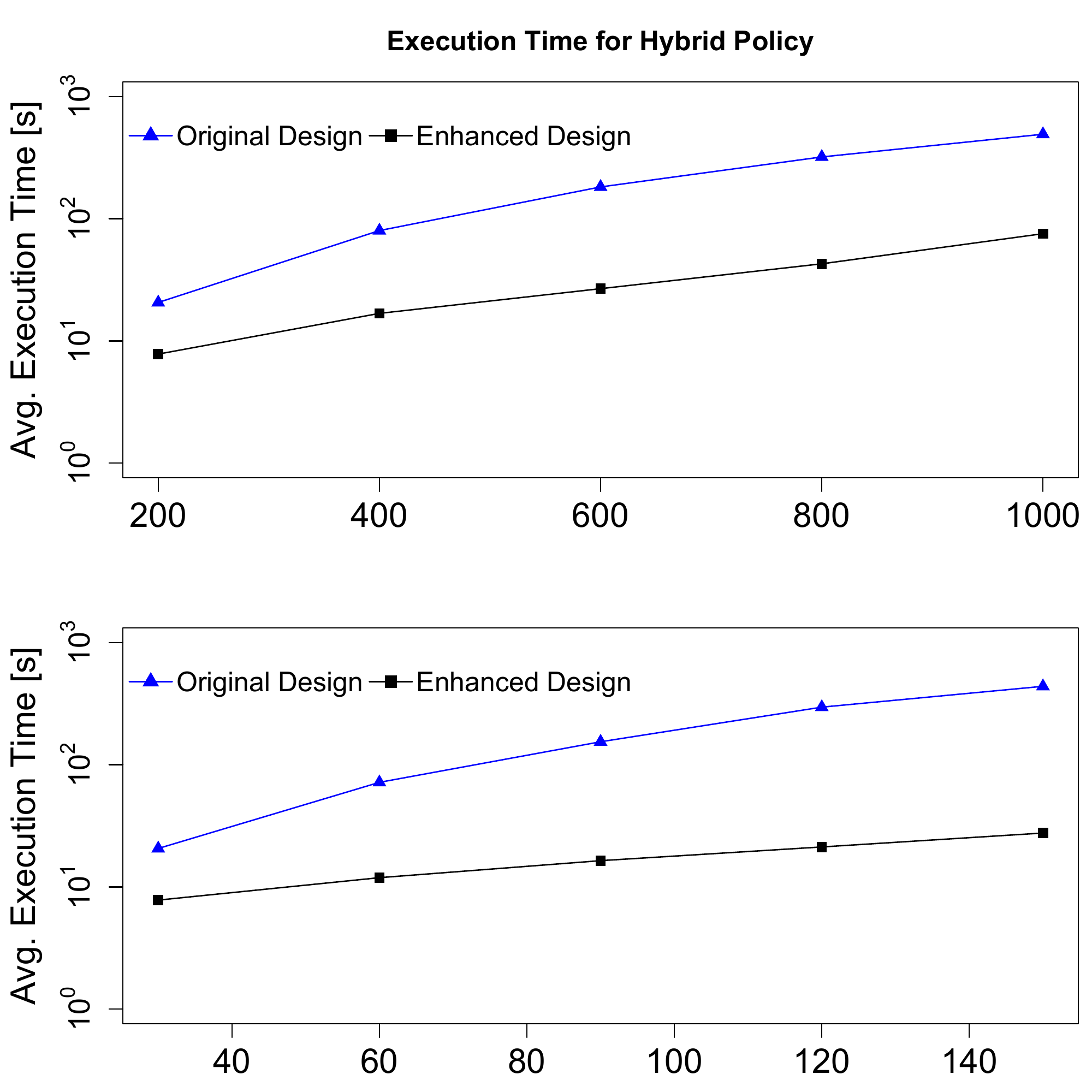}
                \caption{Execution time for hybrid policy of sample app 3 (Y-axis is in log-scale).}
                \label{TGS3HY}
        \end{center}
        \vspace{-0.2in}
\end{figure}

Figure~\ref{TGS1TTEO} shows the results of measured time for sample app 1. The left sub-figure shows the average execution times for the two-tier architecture with edge orchestrator under a fixed simulation duration (i.e., 30 minutes) for different number of devices. The right sub-figure shows the average execution times for different simulation duration. Both sub-figures show that the enhanced simulator significantly outperforms the original design with respect to the average execution time. In addition, when the number of the devices or the simulation duration increases, the gap for the difference of performance increases as well. 
Figure~\ref{TGS2HY} presents the results for sample app 2 for the hybrid edge orchestrator policy.
Both sub-figures show a similar trend as Figure~\ref{TGS1TTEO}, i.e., our enhanced simulator significantly outperforms the original design with respect to the average execution time, and the advantage increases greatly with the increasing of the number of devices or simulation duration.
Lastly, Figure \ref{TGS3HY} illustrates the timing results for sample app 3, where the hybrid edge orchestrator policy was simulated, which allows computation on edge servers as well as on the mobile devices directly. 
Similarly, the enhanced design dominates the original design with respect to the average execution time. The results also show that the enhanced design has better scalability for the number of devices and the simulation duration.

\section{Conclusion}
Because of the growing popularity of the Internet of Things, edge computing concept has been widely studied to relieve the load on the conventional cloud and networks while improving the service quality for end-users. Since experimenting with real infrastructure is often uneconomical or not practical during researches, a discrete-event simulator namely EdgeCloudSim was widely used. In this paper, we enhance several modules in 
the original design without sacrificing any simulation precision. The proposed enhancement not only improves the runtime efficiency of simulation, but also improves the flexibility by fixing the mismatches on the concept of discrete-event simulation. Through extensive experiments, we show that the enhancement does not affect the expressiveness of simulations while obtaining 2 orders of magnitude speedup on average. In future work, we plan to replace all floating-points with integers and introduce more real-time task models.
\section*{Acknowledgement}
\vspace{-0.1in}
This work is partly supported by Deutsche Forschungsgemeinschaft (DFG) within the Collaborative Research Center SFB 876, project A1 and A3 (\url{https://sfb876.tu-dortmund.de}).

\footnotesize


\bibliographystyle{abbrv}

\bibliography{lit}

\begin{thebibliography}{10}

\bibitem{CS:2010}
R.~N. Calheiros, R.~Ranjan, A.~Beloglazov, C.~A.~F. De~Rose, and R.~Buyya.
\newblock Cloudsim: a toolkit for modeling and simulation of cloud computing
  environments and evaluation of resource provisioning algorithms.
\newblock {\em Software: Practice and Experience}, 41(1):23--50, 2011.

\bibitem{CASADEI2019154}
R.~Casadei, G.~Fortino, D.~Pianini, W.~Russo, C.~Savaglio, and M.~Viroli.
\newblock A development approach for collective opportunistic edge-of-things
  services.
\newblock {\em Information Sciences}, 498:154--169, 2019.

\bibitem{encstat}
Y.~Dodge.
\newblock {\em The Concise Encyclopedia of Statistics}.
\newblock Springer New York, New York, NY, 2008.

\bibitem{EdgeCloudSim-imp}
R.~Freymann and K.-H. Chen.
\newblock {The Repository for Light-Weight Design for Edge Computing
  Simulation}.
\newblock \url{https://tu-dortmund.sciebo.de/s/u9KzqDM1pDUNLoI}, 2021.

\bibitem{iFogSim}
H.~Gupta, A.~Vahid~Dastjerdi, S.~K. Ghosh, and R.~Buyya.
\newblock ifogsim: A toolkit for modeling and simulation of resource management
  techniques in the internet of things, edge and fog computing environments.
\newblock {\em Software: Practice and Experience}, 47(9):1275--1296, 2017.

\bibitem{Law2000}
A.~M. Law and W.~D. Kelton.
\newblock {\em Simulation modeling and analysis}.
\newblock New York (N.Y.) : McGraw-Hill, 3rd ed. edition, 2000.

\bibitem{app8071160}
J.~Lee and J.~Lee.
\newblock Hierarchical mobile edge computing architecture based on context
  awareness.
\newblock {\em Applied Sciences}, 8(7), 2018.

\bibitem{EdgeComp}
W.~{Shi}, J.~{Cao}, Q.~{Zhang}, Y.~{Li}, and L.~{Xu}.
\newblock Edge {C}omputing: {V}ision and {C}hallenges.
\newblock {\em IEEE Internet of Things Journal}, 3(5):637--646, 2016.

\bibitem{ECS1}
C.~{Sonmez}, A.~{Ozgovde}, and C.~{Ersoy}.
\newblock Edgecloudsim: An environment for performance evaluation of {E}dge
  {C}omputing systems.
\newblock In {\em 2017 Second International Conference on Fog and Mobile Edge
  Computing (FMEC)}, pages 39--44, 2017.

\bibitem{ECS2}
C.~Sonmez, A.~Ozgovde, and C.~Ersoy.
\newblock Edgecloudsim: An environment for performance evaluation of edge
  computing systems.
\newblock {\em Transactions on {E}merging {T}elecommunications {T}echnologies},
  29(11):e3493, 2018.

\bibitem{DBLP:conf/ccgrid/TomaWLC19}
A.~Toma, J.~Wenner, J.~E. Lenssen, and J.~Chen.
\newblock Adaptive quality optimization of computer vision tasks in
  resource-constrained devices using edge computing.
\newblock In {\em 19th {IEEE/ACM} International Symposium on Cluster, Cloud and
  Grid Computing, {CCGRID} 2019, Larnaca, Cyprus, May 14-17, 2019}, pages
  469--477. {IEEE}, 2019.

\bibitem{QQ}
M.~B. Wilk and R.~Gnanadesikan.
\newblock {Probability plotting methods for the analysis for the analysis of
  data}.
\newblock {\em Biometrika}, 55(1):1--17, 03 1968.

\bibitem{8449105}
Q.~{Zhang}, M.~{Lin}, L.~T. {Yang}, Z.~{Chen}, S.~U. {Khan}, and P.~{Li}.
\newblock A double deep q-learning model for energy-efficient edge scheduling.
\newblock {\em IEEE Transactions on Services Computing}, 12(5):739--749, 2019.

\end{thebibliography}

\end{document}